\documentclass[twocolumn,showpacs,amsmath,amssymb,pra,superscriptaddress]{revtex4-2}

\usepackage{graphicx}
\usepackage{dcolumn}
\usepackage{bm}
\usepackage{subfigure}
\usepackage{blindtext}

\begin{document}

\title{Energy barriers of Be and B in passing through the C$_{60}$ fullerene cage}

\author{A. V. Bibikov}

\affiliation{Skobeltsyn Institute of Nuclear Physics Lomonosov
Moscow State University, 119991 Moscow, Russia}


\author{A. V. Nikolaev} 

\affiliation{Skobeltsyn Institute of Nuclear Physics Lomonosov
Moscow State University, 119991 Moscow, Russia}



\author{P. V. Borisyuk}

\affiliation{National Research Nuclear University MEPhI, 115409,
Kashirskoe shosse 31, Moscow, Russia}

\author{E.~V.~Tkalya}

\affiliation{P. N. Lebedev Physical Institute of the Russian
Academy of Sciences, 119991, 53 Leninskiy pr., Moscow, Russia}


\affiliation{Nuclear Safety Institute of RAS, Bol'shaya Tulskaya
52, Moscow 115191, Russia}


\begin{abstract}
We have studied the potential barriers for the penetration of atomic beryllium or boron inside the C$_{60}$ fullerene by performing {\it ab initio} density functional theory (DFT) calculations with three variants for the exchange and correlation: B3LYP (hybrid functional), PW91 and PBE. Four principal trajectories to the inner part of C$_{60}$ for the penetrating atom have been considered: through the center of six-member-carbon ring (hexagon), five-member-carbon ring (pentagon), and also through the center of the double C--C bond (D-bond) and the center of the single C--C bond (S-bond). Averaging over the three DFT variants yields the following barriers for beryllium penetrating inside a deformable fullerene: 3.2~eV (hexagon), 4.8~eV (S-bond), 5.3~eV (D-bond), 5.9~eV (pentagon). These barriers correspond to the slow and adiabatic penetration of Be, in contrast to the fast (non-adiabatic) penetration through the rigid cage of C$_{60}$ resulting in 5.6~eV (hexagon), 16.3~eV (pentagon), 81.8~eV (S-bond) and 93.4~eV (D-bond). The potential barriers for the boron penetrating inside deformable/rigid C$_{60}$ are: 3.7/105.4~eV (D-bond), 4.0/86.8~eV (S-bond), 4.7/7.8~eV (hexagon), 6.8/14.0~eV (pentagon). The potential barriers for Be and B escaping from the inner part of C$_{60}$ are higher by the value of $\sim$0.84~eV for Be and $\sim$0.81~eV for B. The considerable reduction of the potential barriers for the deformable fullerene is ascribed to the formation of the Be--C and B--C bonds. We discuss the difference between Be and B, compare three variants of DFT, and analyze the role of the dispersion interaction.
\end{abstract}


\maketitle

\section{Introduction}
\label{sec:int}

Novel carbon nanomaterials such as graphene \cite{Nov04,appl} and endofullerenes \cite{Klaus,popov,Shino,Pop17} continue to attract much attention of researchers from
various scientific fields mainly for two reasons.
The first is related to fundamental aspects of their characteristics, while the second is certainly
driven by numerous possible applications in physics, chemistry, material science and biology \cite{medic,Lu}.

One of the discussed application is to use boron-containing endofullerenes for the boron neutron capture therapy (BNCT) \cite{BNCT} in oncology treatment \cite{locher1936,farr1954,BARTH2020}.
In BNCT boronated agents deliver boron-10 to tumors, which, after undergoing irradiation with neutrons, yields lithium-7 and an alpha particle
in accordance with the nuclear reaction $^{10}$B$(n,\alpha)^{7}$Li.
The alpha particle having a short mean free path, preferentially affects tumor tissues and spares more remote normal tissues.
To date, there are only two drugs that currently are used clinically,
sodium borocaptate (BSH) and boronophenylalanine (BPA) \cite{BARTH2020},
which are not free from certain drawbacks.
Using a fullerene cage with one or several boron atoms inside,
may be an interesting option for new radio-pharmaceuticals for BNCT.

On the other hand, the interaction of $^{10}$B with protons of the energy exceeding 100 eV gives rise to the
nuclear reaction $^{10}$B($p$,$\alpha$)$^{7}$Be resulting in the appearance of $^{7}$Be.
The $^{7}$Be nucleus as well as $^{10}$B has a very large cross-section of the interaction with slow neutrons \cite{Rauscher-96},
and therefore, it can also be used for the cancer neutron capturing therapy \cite{Scorciapino-21}.
Since beryllium is a very toxic chemical, the C$_{60}$ fullerene is a necessary element of the molecular structure giving a natural protecting shell for the delivery of $^{10}$B to tumor sites.
Thus, the production of $^{7}$Be in the C$_{60}$ interior region \cite{Ohtsuki-04,Tkalya-10,Bibikov-13,Bibikov-19}
from non-radioactive $^{10}$B can be used in medicine and the endofullerenes $^{7}$Be@C$_{60}$, $^{10}$B@C$_{60}$ \cite{Bibikov-22} should be considered
as prospective nanomaterials.

There are different experimental technics for encapsulation of atoms inside fullerenes \cite{popov}, one of which
is the ion (atom) bombardment, when external atoms are implanted through the walls of the already existing carbon cages
\cite{popov,Tell-96,Camp-97,Murpy-96,Diet-99,Aoy-10}.
Another involves implantation into fullerenes in the gas phase under single-collision conditions \cite{Camp,Saun}.
These experiments have been performed with inert gas atoms or ions, small molecules, alkali ions and transition metal ions,
see for a review Ref.\ \cite{Deng}.
Having carried out, the experiments provide direct information about threshold energies for such atoms, ions and molecules, Table I of \cite{Deng},
which now can be compared with theoretical values.
Theoretical description started with a discussion on possible mechanism of He insertion into C$_{60}$ \cite{Murry,Hru,Cui,Patch}
using both {\it ab initio} calculations \cite{Murry,Hru,Patch} and molecular dynamics \cite{Cui}.
Initially, a specific ``window" mechanism involving breaking of a single C-C bond was suggested \cite{Murry},
with the formation of a large hole in C$_{60}$, through which atoms can pass.
In subsequent studies \cite{Hru,Patch} it became clear that there are various paths leading inside the fullerene, through
the five-membered ring (pentagon), six-membered ring (hexagon) or others \cite{Patch}.
It turned out that for helium the optimal path is through the hexagon, although even in this case the calculated barrier
($9-11$~eV) is higher than experimental values ($\geq$3~eV) \cite{Deng}.
On the other hand, for the nitrogen atom
encapsulated in C$_{60}$ the situation is completely different: according to the semiempirical (PM3-UHF) calculations \cite{Maus,Waib}, lowest potential barriers correspond to the trajectory of the nitrogen atom passing through a middle point of the double (1.8~eV \cite{Maus}, 2.7~eV \cite{Waib}) or single C-C bonds (2.1~eV \cite{Maus}).
That has been explained by the fact that the moving nitrogen atom being close to a C-C bridge of C$_{60}$ forms N-C covalent bonds with the carbon atoms,
which shifts two neighboring carbon atoms on its way farther apart.
The local structure around N is then substantially deformed while the potential barrier is
lowered. As a result, the atom penetrates the fullerene cage through the C-C bond more easy than through the hexagon center \cite{Maus,Waib}.
Such mechanism is not possible for the atom of helium, whose electron shells are inert.
It seems that the main precondition for this path to occur, is a possibility for a penetrating atom to form strong bonds with the nearby carbon atoms,
assisting its way inside the fullerene cage. This scenario is seemingly confirmed by experimental estimations of the activation energy (1.74~eV \cite{Maus} and 1.57~eV \cite{Waib}),
obtained with the Arrhenius equation for the ESR data.

In the present study we investigate the penetration of the atoms of boron and beryllium by modelling
various paths to (or from) the exterior region of the fullerene C$_{60}$ at the {\it ab initio} level.
Both boron and beryllium having valence electrons can form stable molecular complexes with carbon atoms.
Therefore, the intriguing question is which path -- through the center of the hexagon (as for helium) or through
the C-C bonds (as for nitrogen) -- is preferable.
We also pay attention to the energetics of the process, considering the deformation of the C$_{60}$ cage and emerging potential barriers.
To the best of our acknowledge, there is only a single study of penetration of $^7$Be into the C$_{60}$ fullerene by the
recoil implantation following nuclear reaction \cite{Oht}.
A subsequent molecular dynamics simulation has been used to show that a $^7$Be$^{2+}$ ion with 5~eV kinetic energy
can pass trough the center of the six-membered ring \cite{Oht}.


The paper is written as follows: in Sec.~II we briefly describe our method, the variants of applied DFT functionals (B3LYP, PW91, PBE) and other details of calculations.
The results are presented in Sec.\ \ref{sec:res}, in subsection \ref{Be} - for beryllium, in subsection \ref{Boron} - for boron.
In Sec.~\ref{DFT-comp} we compare results obtained with different variants of the exchange-correlation contribution.
Our main findings and conclusions are summarized in Sec.~\ref{sec:con}.

\section{Method}
\label{sec:method}

The calculations have been performed within the {\it ab initio} density functional theory (DFT) approach \cite{QC,DFT} as
implemented in Ref.\ \cite{GAMESS}.
The adopted molecular basis sets were 6-31G$^{**}$ for carbon, beryllium and boron.

Within the DFT approach, we have used the B3LYP hybrid functionals \cite{B3LYP-1,B3LYP-2}
mixing the DFT exchange with the pure Hartree-Fock one for the resulting exchange-correlation potential.
The dispersion interaction has been included in calculations by adopting the semi-empirical Grimme DFT-D3 method \cite{Grimme3}.
The performance of the B3LYP hybrid functional is compared with
results obtained within a more simple (non-hybrid) Perdew-Burke-Ernzerhof (PBE) exchange-correlation model \cite{PBE},
and also, commonly used the Perdew-Wang 1991 exchange correlation functional (PW91) \cite{PW91}.

The C$_{60}$ molecule has a very high icosahedral ($I_h$) symmetry.
It has 20 six-carbon-rings (hexagons) and 12 five-carbon-rings (pentagons), with so called double and single C--C bonds, which are not equivalent.
Thirty double bonds, often marked as [6,6] ones, fuse two neighboring hexagons, and 60 single bonds -- or [5,6] ones -- fuse neighboring pentagons and hexagons.
Double bonds are stronger and have smaller C-C bond length ($l_{db} < l_{sb}$).
Experimentally, $l_{db}=1.400(15)$~{\AA}, $l_{sb}=1.450(15)$~{\AA}
from $^{13}$C nuclear magnetic resonance ($^{13}$C NMR) measurements \cite{C60-NMR}, and
$l_{db}=1.404(10)$~{\AA}, $l_{sb}=1.448(10)$~{\AA} from high resolution neutral scattering \cite{C60-neut},
with the molecular radius $R_{C60}=3.55$~{\AA}.
Our calculations with the 6-31G** basis set with the B3LYP functional yield $l_{db}=1.396$~{\AA}, $l_{sb}=1.454$~{\AA}, and $R_{C60}$=3.551~{\AA}.

In analyzing the barriers, as the zero level of energy we adopted the energy of the separated distant atom (B or Be) and the C$_{60}$ fullerene.
Therefore, the energy at close distances coincides with the binding energy for the molecular complex BC$_{60}$ or BeC$_{60}$,
\begin{eqnarray}
    E_b = E(\textrm{A}\textrm{C}_{60}) - E(\textrm{C}_{60}) - E(\textrm{A}) ,
    \label{m1}
\end{eqnarray}
where A stands for the atom of B or Be with the energy $E(\textrm{A})$, and $E(\textrm{C}_{60})$ is the energy of the undistorted isolated C$_{60}$ fullerene.
The potential barriers are identified as maxima of the energy $E_b$, which occur when the penetrating atom enters the fullerene cage.

Further, we assume that carbon atoms of the fullerene follow adiabatically the movement of the penetrating atom (B or Be), i.e. the fullerene vibrations
fully assist to B (or Be) in getting inside the fullerene.
Although it is not discussed explicitly, this approximation has been adopted practically in all earlier studies \cite{Patch,Maus,Waib}.
The geometry optimization has involved the two carbon shells of C$_{60}$ nearest to the penetrating atom, which amounts to 12 carbon atoms for
the penetration through the center of hexagon, 10 carbon atoms for the path through the center of pentagon,
14 and 13 carbon atoms for the path through the double and single bond midpoint, respectively.
At the first stage depending on the chosen trajectory, the $C_{5v}$, $C_{3v}$, or $C_{2v}$ rotation symmetry was imposed.
The final relaxation of the carbon cage at various fixed positions of the beryllium or boron atom has been performed
at a very low symmetry level ($C_s$) with only one vertical mirror plane.

The center of the fullerene ($\vec{R}_c$) can be considered fixed because only the carbon atoms facing the incoming atom of Be or B were allowed to be displaced, while the carbon atoms on the opposite side of C$_{60}$ and the center, defined in respect to them, were kept fixed. We then characterized the position of the penetrating atom $A$ by
the radius vector $\vec{R} = \vec{R}_A - \vec{R}_c$ and performed DFT calculations of the binding energy $E_b(R)$ as a function of $R$, Eq.\ (\ref{m1}),
for a set of several radial points $R_i$ along chosen directions of $\vec{R}$,
defined in Sec.\ \ref{sec:res} below.
At the radii corresponding to the maxima of $E_b(R)$ we carried out calculations aiming at obtaining the transition states.
When such calculations were successful (e.g. for the penetration through the hexagon) the potential barrier was defined with the help of the transition state. However, quite often we could not find the transition state for the reason discussed in detail in the subsection \ref{Shape} below.
In such cases we defined the potential barrier using the maximal value of $E_b(R_i)$ along the chosen trajectory and denoted the corresponding radial value as $R_v$.

In the study we also discuss the barriers for rigid (non-deformed) C$_{60}$, when all carbon atoms do not change their positions
in response to the penetrating atom. This is the opposite case to the adiabatic penetration of atom.

\section{Results and discussion}
\label{sec:res}

\begin{figure}
\resizebox{0.45\textwidth}{!} {
\includegraphics{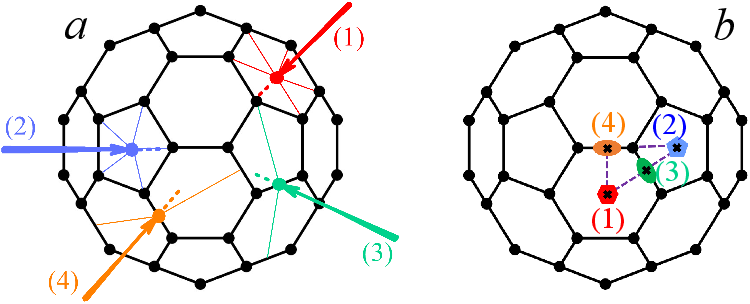}}
\caption{
Four trajectories for the penetration of an atom (Be or B) inside the C$_{60}$ fullerene shown by red (1), blue (2), orange (3), and green (4) line,
panels $a$ and $b$:
(1) through the center of the six-member-carbon ring (hexagon), (2) through the center of the five-member-carbon ring (pentagon),
(3) through the center of the single C--C bond (S-bond), and (4) through the center of the double C--C bond (D-bond).
$b$ -- irreducible part of C$_{60}$ for the penetration.
} \label{fig1}
\end{figure}

Considering possible trajectories of the penetrating atom (B or Be), leading to the inner region of the fullerene cage,
we easily distinguish two simple paths: through the hexagon (1) and through the pentagon (2), Fig.~\ref{fig1}.
Intuitively, these paths are expected to have lower potential barriers.
This turns out to be true, but only for the undeformable (rigid) fullerene, Table~\ref{tab1}.
However, there are also two other, less obvious, ways of getting inside -- through the single C-C bond (3) and through the double C-C bond (4), Fig.~\ref{fig1}.
For the frozen fullerene geometry in the last two cases (3,4) the atomic repulsion clearly prevails, making them highly unfavorable
energetically, Table~\ref{tab1}. The situation changes drastically for a deformable molecule, when carbon atoms in response to the penetrating atom are displaced
forming local chemical bonds with it.
This leads to a large lowering of the total energy and, as a result, to small energy barriers for the penetrating atom, Table \ref{tab2}.
%
%

In the following we limit ourselves to these four main types of trajectories. 
We also assume that the paths through hexagon and pentagon
go through their centers along the radial fullerene direction.
Analogously, the radial path through the single and double bond passes through the middle bond point.

It is important to realize that the four trajectories form an irreducible sector for a penetration through
the fullerene, shown in Fig.~\ref{fig1}b. The trajectories through the hexagon (1), pentagon (2) and D-bond (4), form
vertices on the surface of the C$_{60}$ molecule, Fig.~\ref{fig1}b, and any path leading inside can be reduced to
the one passing through a point in the surface area bordered by the lines connecting vertices (1)-(3)-(2)-(4)-(1).
Therefore, the trajectories from (1) to (4) being critical, Fig.~\ref{fig1}b, are highly representative, providing with
potentially extremal paths inside the fullerene.

%
\begin{table}[hbt!]
\caption{
The energy barriers for the penetration of Be or B inside the rigid (undeformed) C$_{60}$ fullerene (in eV) through different paths,
calculated with the B3LYP, PW91 and PBE exchange-correlation potentials.
Two variants -- with the dispersion (van der Waals) interaction (vdW) and without it -- are given.
(Two values marked by star ($^*$) are with $S=1$.)
\label{tab1} }
\begin{ruledtabular}
\begin{tabular}{c  l  c  c  c  c  c}

 DFT & & path: & hexagon & pentagon & D-bond & S-bond \\
\colrule
           & Be  & vdW & 8.16 &  16.57   & 94.15  & 82.42 \\
 B3LYP     & Be  &     & 8.25 &  16.67   & 94.19  & 82.47 \\
           & B   & vdW & 8.73 & 14.92 & 105.75 & 87.11 \\
           & B   &     & 8.80 & 15.00 & 105.79 & 87.16 \\
\hline
           & Be      & vdW & 4.40 & 16.08 & 92.85 & 81.31 \\
 PW91      & Be      &     & 4.49 & 16.19 & 92.90 & 81.38 \\
           & B       & vdW & 7.36 & 13.50 & 105.04 & 86.50 \\
           & B       &     & 7.43 & 13.58 & 105.08 & 86.56 \\
\hline
           & Be      & vdW & 4.16$^*$ & 16.30 & 93.21 & 81.65 \\
 PBE       & Be      &     & 4.20$^*$ & 16.35 & 93.24 & 81.69 \\
           & B       & vdW & 7.39 & 13.60 & 105.46 & 86.91 \\
           & B       &     & 7.41 & 13.63 & 105.48 & 86.93 \\
\end{tabular}
\end{ruledtabular}
\end{table}

\subsection{Energy barriers for beryllium}
\label{Be}

%
\begin{table}[hbt!]
\caption{
The potential energy barriers for the penetration of Be inside the C$_{60}$ fullerene (in eV) through different paths,
calculated with various variants of DFT exchange-correlation potential.
D-bond stands for the penetration through the double bond center, S-bond through the single bond center.
Two variants -- with the dispersion (van der Waals) interaction (vdW) and without it -- are listed.
$R_v$ is the radius with the largest path energy.
\label{tab2} }
\begin{ruledtabular}
\begin{tabular}{l  c  c  c  c  c  c  c}

 path     & DFT type: & \multicolumn{2}{c}{B3LYP}       &  \multicolumn{2}{c}{PW91}        & \multicolumn{2}{c}{PBE} \\
 through  & $R_v$     & vdW &       &  vdW  &       &  vdW  &       \\
\tableline
 hexagon  & 3.46 & 3.82 &	3.87 &	2.84 &	2.90 &	2.87 &	2.90 \\
 pentagon & 3.58 & 6.55 &	6.67 &	5.56 &	5.65 &	5.66 &	5.70 \\
 D-bond   & 3.26 & 5.78 &	5.82 &	5.04 &	5.08 &	5.09 &	5.12 \\
 S-bond   & 3.17 & 4.98 &   5.05 &  4.65 &  4.71 &  4.72 &  4.75 \\
\end{tabular}
\end{ruledtabular}
\end{table}

\begin{figure}
\resizebox{0.45\textwidth}{!} {
\includegraphics{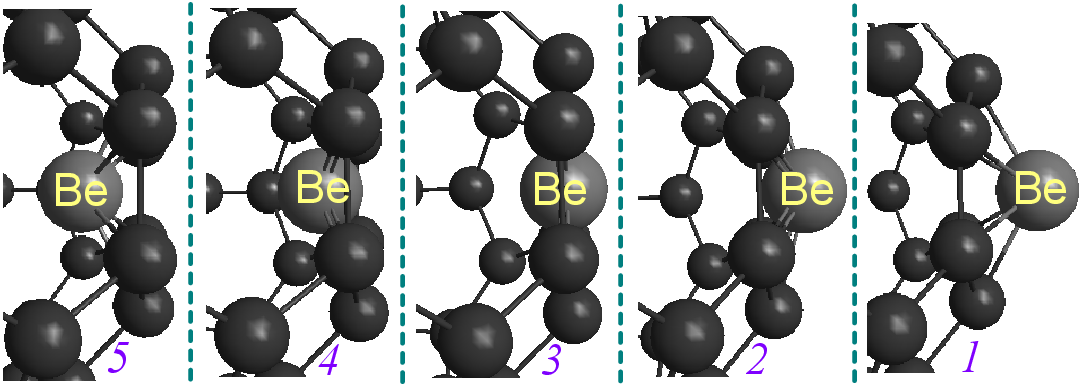}}
\caption{
The penetration of Be through the center of the deformable hexagon with the lowest energy barrier (B3LYP).
Different stages of the penetration of Be are marked by labels 1--5:
(1) $R=4.2$; (2) 3.8; (3) 3.4; (4) 3.0; (5) 2.6, in {\AA}.
In the hexagon opening the double and single C-C bond lengths are (1) 1.45/1.48; (2) 1.50/1.55;
(3) 1.55/1.64; (4) 1.54/1.60; (5) 1.50/1.53, in {\AA}.
} \label{fig2}
\end{figure}

\begin{figure}
\resizebox{0.48\textwidth}{!} {
\includegraphics{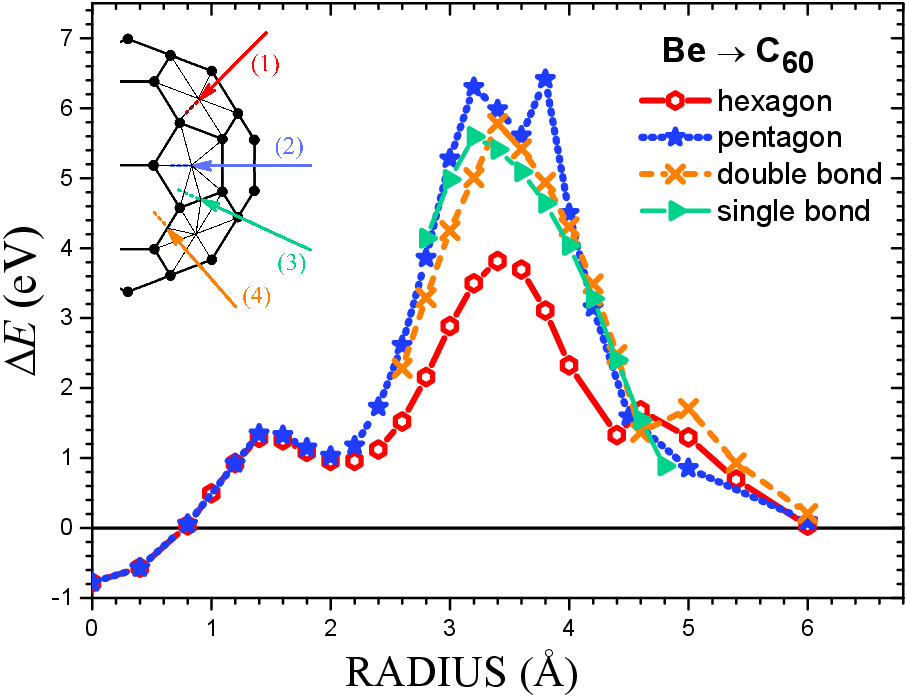}}
\caption{
Binding energies $E_b$ as functions of the penetration radius $R$ of Be
for four different paths through the deformable cage of C$_{60}$ (B3LYP).
The maximum of each $E_b(R)$ gives the corresponding potential barrier.
} \label{fig3}
\end{figure}

The energy barriers for various paths of Be leading inside the undeformable C$_{60}$ are given in Table \ref{tab1}.
[We recall that the zero energy corresponds to Be and the fullerene situated far away from each other, Eq.~(\ref{m1}).]
The inspection of Table \ref{tab1} shows that the easiest way inside the fullerene is through the center of the hexagon
carbon ring.
This conclusion holds for both rigid and deformable C$_{60}$, although in the deformable case
the potential barrier becomes 2.1--2.9 times smaller.
The adiabatic penetration of Be through the deformable hexagon of C$_{60}$ is visualized in Fig.~\ref{fig2}.
The largest potential barriers for the rigid fullerene are found for trajectories through the single and double bonds.
For a deformable molecule, however, the double and single bond barriers drop by more than 10 times and lie very close to each other,
orange ($\times$) and green ($\triangleright$) plots in Fig.~\ref{fig3}.
As a result, the barrier through the single bond is only 1.8--2.2 eV higher than through the hexagon, in sharp contrast with the
rigid fullerene where the energy difference exceeds 70 eV.
In the deformable case the largest barrier is found for the penetration through the pentagon, blue ($\star$) plot in Fig.~\ref{fig3}.
The dependence of the binding energies $E_b$ on the radius $R$ along the four paths of Be, leading inside the deformable fullerene,
from which the adiabatic potential barriers are obtained, are exemplified in Fig.~\ref{fig3} for the B3LYP case.
As shown in Fig.~\ref{fig3}, the peaks for the double and single bond have cusp-like structure (with discontinuous first derivative on $R$)
whereas for the pentagon two peaks are visible.
These peculiarities are related to the change of the electron ground state configuration and discussed in detail in Sec.\ \ref{Shape} below.
The minima developed at around 2 {\AA} in Fig.~\ref{fig3} are due to the formation of chemical bonds between Be and neighboring carbon atoms
of C$_{60}$. Such intermediate chemical bondings are often found for various encapsulated atoms close to the fullerene cage, e.g. \cite{Bibikov-22}.

%
%
%
%

For the penetration through the hexagon (or pentagon) the maximal energy corresponds to approximately planar arrangements of beryllium with six (five) nearest carbon atoms.
For the case of hexagon this stage is shown in panel 3 of Fig.~\ref{fig2}. At this stage the Be--C bonds, which start to form earlier (already at the stage 1 of Fig.~\ref{fig2}) are the strongest with the smallest averaged bond length $d$(Be--C)=1.58~{\AA}, whereas the C--C double and single bonds in the hexagon around Be are the weakest.
The averaged C--C-bond lengths in the hexagon are  1.55~{\AA} and 1.64~{\AA} for the former C--C double and single bond, respectively, which is 11-13\% larger than the values in the isolated fullerene. It is also worth noting that the minimum of energy is reached for the broken C$_3$ hexagon symmetry although this gives
a rather small energy gain of a few hundredths of eV.
In subsequent stages 4 and 5 in Fig.~\ref{fig2} the Be--C bond lengths increase while C--C bond lengths in the hexagon decrease. Thus, the Be--C bonds are weakened and the C--C double and single bonds regain their strength and the hexagon initial geometry becomes recovered.

For trajectories through the single or double bond two neighboring carbon atoms are pushed aside,
and the local fullerene structure around beryllium is strongly deformed -- we consider this effect in detail in Sec.~\ref{Boron} below.
The deformation of the C$_{60}$ cage for various paths of penetration is also shown in Supplemental Material.

%
\begin{table}[hbt!]
\caption{
The binding energy $E_c$ for Be and B located at the center ($R=0$) of the C$_{60}$ fullerene (in eV),
calculated with various variants of DFT exchange-correlation potential.
Two variants -- with the dispersion (van der Waals) interaction (vdW) and without it -- are given.
\label{tab3} }
\begin{ruledtabular}
\begin{tabular}{l  c  c  c  c  c  c  c}

      DFT type: & \multicolumn{2}{c}{B3LYP}       &  \multicolumn{2}{c}{PW91}        & \multicolumn{2}{c}{PBE} \\
                & vdW &       &  vdW  &       &  vdW  &       \\
\tableline
 Be@C$_{60}$   & -0.78 & 0.59 & -1.28 & 0.09 & -0.46 & 0.12 \\
 B@C$_{60}$  &	-0.79 & 0.28 & -1.14 & -0.07 & -0.50 & -0.05 \\
\end{tabular}
\end{ruledtabular}
\end{table}
The optimal position of Be inside the fullerene is at center of the C$_{60}$ fullerene, when $R=0$ \cite{Tkalya-10}.
In the absence of the dispersion forces the binding energy of this BeC$_{60}$ endofullerene is positive, Table~\ref{tab3}, i.e. $E_c = E_b(R=0) > 0$.
The positive sign of $E_c$ implies that the configuration with a remote atom of Be outside the fullerene lies lower in energy ($E_b = 0$).
The inclusion of van der Waals interaction, whose contribution is the largest for the $R=0$ position of Be, however, changes the situation and the location of Be
at the center of C$_{60}$ becomes more favorable, with the energy gain $E_c$ ranging from -0.46~eV (PBE) to -1.28~eV (PW91), Table~\ref{tab3}.
Therefore, the potential barriers for escaping from inside ($\triangle E_0$)
are higher than for penetrating in the inner region of the fullerene from outside ($\triangle E_{\infty}$), quoted in Tables \ref{tab1}, \ref{tab2}.
In general, $\triangle E_0$ can be calculated from $\triangle E_{\infty}$ according to
\begin{eqnarray}
    \triangle E_0 = \triangle E_{\infty} - E_c .
    \label{Be1}
\end{eqnarray}
In all other aspects the inclusion of dispersion forces has only marginal effect on barriers, as follows from calculated values
presented in Tables \ref{tab1}, \ref{tab2}. This also holds for the case of boron, considered in the next section.

\subsection{Energy barriers for boron}
\label{Boron}

For the boron atom, penetrating in the inner part of the rigid cage of the C$_{60}$ fullerene, we find that, as for beryllium, Sec.~\ref{Be},
the easiest trajectories go through the carbon hexagon (7.4--8.7~eV) and pentagon (13.5--14.9~eV), Table~\ref{tab1}.
Paths through the double bond ($\sim105$~eV) and single bond ($\sim87$~eV) requires an order of magnitude more energy.

\begin{figure}
\resizebox{0.45\textwidth}{!} {
\includegraphics{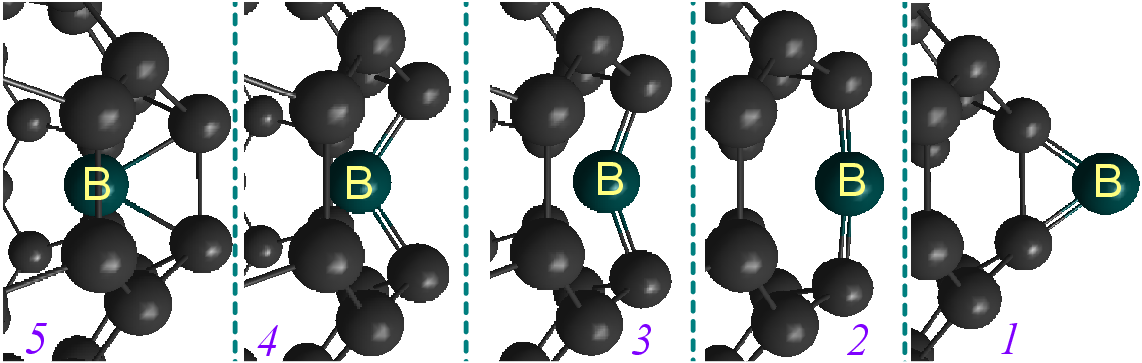}
}
\caption{
The penetration of B through the center of the deformable double bond (D-bond), having the lowest energy barrier (B3LYP).
Different stages of the penetration of B are marked by labels from 1 to 5:
(1) $R=5.0$; (2) 4.0; (3) 3.4; (4) 3.0; (5) 2.6, in {\AA}.
Boron on its move pushes two neighboring carbon atoms aside quite appreciably.
The stretched double C-C bond length is (1) 1.76; (2) 2.70; (3) 2.59; (4) 2.38; (5) 1.56, in {\AA}.
} \label{fig4}
\end{figure}

\begin{figure}
\resizebox{0.48\textwidth}{!} {
\includegraphics{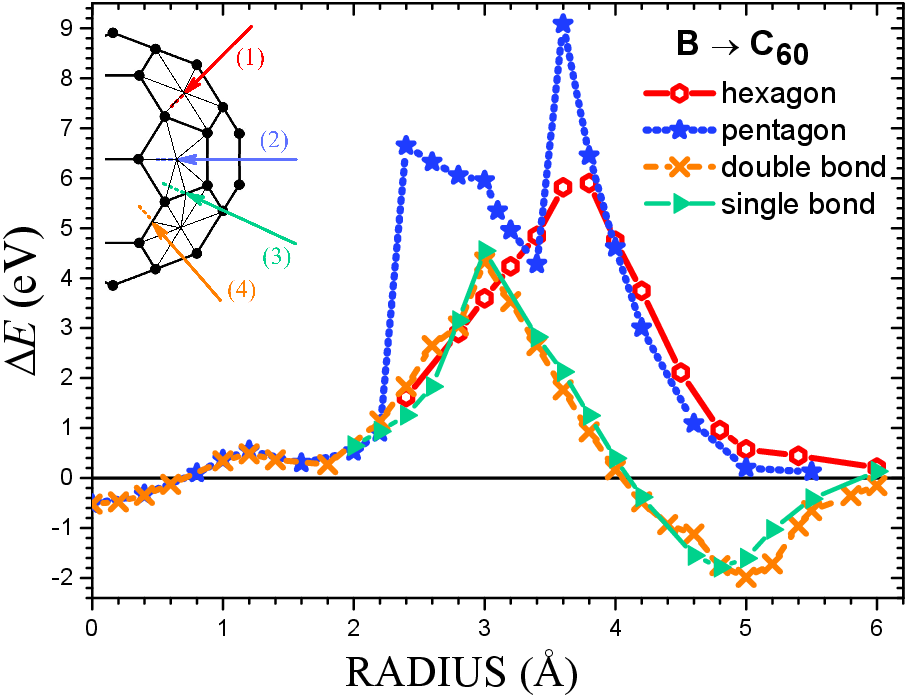}}
\caption{
Binding energies $E_b$ as functions of the penetration radius $R$ of B
for four different paths through the deformable cage of C$_{60}$ (B3LYP).
The maximum of each $E_b(R)$ gives the corresponding potential barrier.
} \label{fig5}
\end{figure}

However, exactly as with Be, the energy barriers decrease substantially if carbon atoms of C$_{60}$ assist the penetration by adopting displacements in accordance with the approaching boron atom, Table~\ref{tab4}.
%
\begin{table}[hbt!]
\caption{
The potential energy barriers for the penetration of B inside the C$_{60}$ fullerene (in eV) through different paths,
calculated with various variants of DFT exchange-correlation potential.
D-bond stands for the penetration through the double bond center, S-bond through the single bond center.
Two variants -- with the dispersion (van der Waals) interaction (vdW) and without it -- are listed.
$R_v$ is the radius with the largest path energy.
\label{tab4} }
\begin{ruledtabular}
\begin{tabular}{l  c  c  c  c  c  c  c}

 path    & DFT type: &   \multicolumn{2}{c}{B3LYP}       &  \multicolumn{2}{c}{PW91}        & \multicolumn{2}{c}{PBE} \\
 through & $R_v$ &  vdW &       &  vdW  &       &  vdW  &       \\
\tableline
 hexagon  & 3.60 &  5.82 &	5.85 &	3.96 &	3.98 &	4.24 &	4.25 \\
 pentagon & 3.80 &  9.08 &	9.17 &	5.57 &	5.60 &	5.83 &	5.86 \\
 D-bond   & 3.00 & 4.32 &	4.38 &	3.44 &	3.51 &	3.48 &	3.51 \\
 S-bond   & 3.00 & 4.55 &	4.61 &	3.64 &	3.68 &	3.69 &	3.72 \\
\end{tabular}
\end{ruledtabular}
\end{table}
The most radical changes accompany the penetrations through the double bond and single bond -- their energies drop 25-30 and 19-24 times, respectively.
As a result the potential barriers through the double bond (3.44-4.32~eV) and single bond (3.64-4.55~eV) become the lowest in energy,
going even below the energy barrier through the carbon hexagon (3.96-5.82~eV), Table~\ref{tab4}.
In comparison with the case of rigid C$_{60}$, Table~\ref{tab1}, this signifies a qualitative change in the order of
optimal penetrating trajectories. In non-deformed fullerene the easiest way to get inside is through the hexagon, while the most energy consuming
is through the double bond. For the deformed C$_{60}$ it is the opposite: the easiest path goes through the C-C double bond whereas
the most difficult path is through the pentagon.
Apparently, this property follows from the fact that boron being more active with carbon atoms than beryllium,
establishes chemical bonds with them more effectively, Fig.~\ref{fig4}.
The dependence of the binding energies $E_b(R)$ on the radius $R$ along the four paths of B with deformable fullerene for the B3LYP functional
is shown in Fig.~\ref{fig5}.
At $\sim 1.6$~{\AA} weak secondary minima are detected for B located inside C$_{60}$. For the outside region,
comparing Fig.~\ref{fig5} with that for Be, Fig.~\ref{fig3}, we find that the paths of boron through the double and single bonds
[the orange ($\times$) and green ($\triangleright$) plots in Fig.~\ref{fig5}] demonstrate a pronounced minimum around 4.9~{\AA}, which is absent in the case of Be.
This position of B is marked as the stage 1 in Fig.~\ref{fig4}.
The appearing of this minimum is directly related to a bound state of the boron atom situated outside the C$_{60}$ fullerene \cite{Bibikov-22}.
The B--C bond length there is $1.52$~{\AA}, whereas the C--C bond length amounts to $1.76$~{\AA}, which is 26{\%} more than in rigid fullerene.
When boron reaches the stage 2 in Fig.~\ref{fig4} ($R=4$~\AA) it is located approximately at the middle point of the former double C--C bond, causing
its stretching to 2.70~{\AA}, which is almost two times larger than in the rigid fullerene. The B--C bond length at this stage is minimal ($\sim1.35$~\AA).
Progressing to the stages 3 to 5 results in a considerable weakening of the B--C bond with subsequent restoring the C--C double bond.

The visible two peak structure of the potential profile for the penetration through the pentagon for B, Fig.~\ref{fig5}, and Be, Fig.~\ref{fig3}, is directly related to a very complex behavior of the electron bonds of the five member carbon ring. This and other peculiarities of the peaks in Fig.\ \ref{fig5} are discussed in more detail in Sec.\ \ref{Shape} below. The deformation of the C$_{60}$ cage for various paths of penetration is shown in Supplemental Material.

Inspection of Table~\ref{tab3} shows that in preferable energy position, when B is located at the center of the fullerene \cite{Vinit,Bibikov-22},
the binding energy $E_c$ of the molecular complex, including the wdW contribution, is negative.
This implies that the penetration barriers $\triangle E_0$ to escape from the center of the fullerene are by the same value of $|E_c|$ larger
than the potential barriers $\triangle E_{\infty}$, quoted in Table~\ref{tab4}, Eq.~(\ref{Be1}).

%
%
%
%

\subsection{Peculiarities of potential energy barriers}
\label{Shape}

In Fig.\ \ref{fig3} for Be and in Fig.\ \ref{fig5} for B (B3LYP variant of DFT) the peaks for the penetration through the double and single bond display cusps with
discontinuous first derivative on $R$ at maxima.
This is closely related to the fact that at these places we observe a change in the electronic structure due to the intersection of the HOMO-LUMO levels.
Consider for example, the passage of B through the double bond of C$_{60}$, where the HOMO and LUMO levels intersect at about 2.8~{\AA} there.
This conclusion is confirmed by the change of the character of the irreducible representation of the HOMO level - from $A_1$ (at 2.78~\AA) to $B_1$ (at 2.85~\AA),
see Supplemental Material for more details.
As a consequence, the contribution from the HOMO state to the total electron density is very different at $R < 2.78$~{\AA} and at $R > 2.85$~{\AA},
i.e. does not change smoothly through the transition. Due to this transformation, the region from 2.78 to 2.85~{\AA} is not accessible
characterizing by the practical absence of convergence. Therefore, in going from 2.78~{\AA} to 2.85~{\AA} the electron density changes irregularly, which results in the cusps of the peaks and a discontinuous change of the first derivative on $R$ through the maximum point. That is the situation with the penetration of B and Be through the double and single bond. For other cases it is different. The penetration
through the hexagon is smooth (with continuous first derivatives) and we were able to find the transition state there.

The visible two peak structure of the potential profile for the passage through the pentagon for B, Fig.~\ref{fig5}, and Be, Fig.~\ref{fig3}, is also explained by
a sudden reconstruction of the ground state.
We relate these two step-wise changes to the pentagon window opening and closing, resulting in very different electron configurations.
Our calculation indicate that as the position of B is changed from 3.60 {\AA} to 3.40 {\AA}
the averaged pentagon radius is sharply increased from 1.21 to 1.61 {\AA}. This is accompanied with two carbon atoms moving closer to B causing
an appreciable deformation of the pentagon. This new type of bonding persists down to $R=2.2$~{\AA} where the pentagon opening get closed,
the averaged pentagon radius is decreased from 1.59 to 1.25 {\AA} and all C-C bonds in pentagon become equal again. (See Supplemental Material for more details.)
Although the window opening also occurs in other cases, for the pentagon it is much more pronounced and very sharp.
Our finding is in line with results of Ref.\ \cite{Mal}, where the authors could not find the transition state for the passage of Li$^+$, Na$^+$, and K$^+$ through the pentagon, and their barrier profiles also display points of discontinuous change of the first derivative, Fig.~3 and Fig.~4 of Ref.\ \cite{Mal}.
We think that in our case the change of the ground state configuration occurs more readily because we allow for the lowering of high 5-fold rotation symmetry
of the pentagon.

\subsection{Comparison of results with different exchange-correlation potentials}
\label{DFT-comp}

In comparing results for potential barriers obtained with different variants of DFT (i.e. B3LYP, PW91 and PBE),
we mention that the values obtained with PW91 and PBE certainly lie more close together than that of B3LYP.
This holds because the Hartree-Fock, present as a contributing term in the hybrid functional B3LYP, systematically gives larger values
for the barriers.
The maximal deviation in the values of barriers between PW91 and PBE is 0.1~eV for Be and 0.28~eV for B, whereas in the case of B3LYP and PW91
it is almost 1~eV for Be and 1.9~eV for B (both for the penetration through hexagon).

The difference in barriers of various variants of DFT is increased due to the deformations of the fullerene cage,
which are also changed in going from B3LYP to PW91 and PBE. For the fixed geometry of Be@C$_{60}$ and B@C$_{60}$, for example,
where in both complexes the atom of Be/B resides at the center of the fullerene, the energy difference is smaller than 1~eV, Table~\ref{tab3}.

Although the data for barriers are often scattered numerically, qualitatively all values of potential barriers within B3LYP, PW91 and PBE follow
the same pattern.
Therefore, for briefness and representativeness
we have decided to introduce the values of barriers averaged over the three variants of the exchange and correlation (i.e. B3LYP, PW91 and PBE),
which we have quote earlier in the abstract.

\section{Conclusions}
\label{sec:con}

We have studied the potential barriers for beryllium and boron penetrating in the inner part of the C$_{60}$ fullerene
by performing DFT calculations with three types of the exchange-correlation functional: B3LYP (hybrid potential), PW91 and PBE.

In our studies we consider two extreme situations: the first corresponds to the completely rigid C$_{60}$ fullerene,
whereas in the second case the carbon atoms of C$_{60}$ change their locations by adapting places with the minimal total energy in respect to
a fixed position of the penetrating atom.
Although in the literature calculations are usually performed for the deformable C$_{60}$ (e.g. \cite{Patch,Maus,Waib}),
these two approximations, in fact, represent the adiabatic versus diabatic dichotomy well known in quantum mechanics. If the penetrating atom (Be or B)
approaches the fullerene `slowly' the carbon atoms of C$_{60}$ have enough time to adapt their location and `assist'
the act of penetration. The energy barriers for this adiabatic process are quoted in Table~\ref{tab2} for Be and in Table~\ref{tab4} for B.
They are several times smaller in comparison with the penetration of the atom through the carbon cage of completely rigid fullerene.
Potential barriers for that fast (non-adiabatic) process of penetration are listed in Table~\ref{tab1}.
Both types of penetration are not forbidden theoretically and presumable they constitute the smallest and the largest values
for corresponding energy barriers.
In this connection it is worth noting that in some experimental studies the potential barriers are rather defined as threshold energies
to underline that they constitute the lowest possible energy of insertion \cite{Deng}.
In some works two distinct thresholds are mentioned pertaining to different insertion mechanisms \cite{Deng}, which fits well
the considered possibilities of different barriers for different paths.

One of the main consequences of allowing carbon displacements of C$_{60}$ is a significant (more than an order of magnitude)
reduction of the barriers for the penetrations through the middle of the C--C double bond and the middle of the C--C single bond.
For boron this effect makes the path through the double bond the lowest in energy (3.4-4.3~eV), although the same type of barrier
for the rigid C$_{60}$ amounts to 105-106~eV.
For the rigid fullerene the smallest barrier for the penetration of the boron atom is through the hexagon (7.4-8.7~eV).
For beryllium the path through the hexagon is preferable for all cases, with the barrier reaching 2.8--3.8~eV for the deformable C$_{60}$
and 4.2--8.2~eV for the undeformable molecule. Other barriers are listed in Tables~\ref{tab2} and \ref{tab4}.

Inside C$_{60}$ both Be and B have the lowest energy at the center of the fullerene ($R=0$).
The binding energy of Be@C$_{60}$ and B@C$_{60}$ (with the vdW contribution) in this geometry is negative ($E_c < 0$), Table~\ref{tab3},
implying that the potential barriers for escaping from inside are by $|E_c|$ larger than for the penetrating the cage of C$_{60}$ from outside.
The energy difference $|E_c|$ between the barriers is 0.5-1.3~eV for Be and 0.5-1.1~eV for B, Table~\ref{tab3}.

Finally, we would like to remark on practical implication of our work.
Experimentally, it has been proven that Be can be encaged in C$_{60}$ by a nuclear recoil implantation technique \cite{Oht}.
This method, however, involves irradiation with protons of high energy (12~MeV).
In this process Be atoms are loosing their kinetic energies by destroying many fullerenes on its way \cite{Oht}.
Our study indicates that the synthesis of Be@C$_{60}$ and B@C$_{60}$ can be reached either by gas-phase collisions between fullerenes and atoms of Be and B
with much lower threshold energy or by bombarding solid C$_{60}$ with these atoms.

Our estimations, based on our potential energy profiles, indicate that
the additional kinetic energy of carbon atoms transferred by opening and closing windows for penetration of Be/B is approximately 1~eV. This amount
signifies an energy loss of Be/B required for assisting its penetration through the C$_{60}$ cage.
Taking this energy loss into account, the optimal initial energy of incident atom before penetration lies from 4.2~eV to 5.2~eV for Be,
and from 4.7~eV to 5.7~eV for B.
Increasing the upper limit results in an enhancement of the probability of
escaping from the inner region of the fullerene. The consideration above refers only to a single event of penetration. If the incident atom (Be/B) has energy higher
than 5.2/5.7~eV, it simply loses 2~eV of its energy in passing through the fullerene (double penetration through the C$_{60}$ cage) and move to
the next fullerene molecule.
It is worth mentioning that N@C$_{60}$ can be obtained by nitrogen implantation in solid C$_{60}$ \cite{Murp} in this way.
Since both nitrogen and boron are neighbors of carbon in the periodic table of elements, the properties of N@C$_{60}$ and B@C$_{60}$
should be similar in this regard, and we expect that the B@C$_{60}$ fullerene can also be obtained by boron implantation in solid C$_{60}$.

%

\acknowledgements

The reported study was funded by a bilateral Russian (RFBR)-Italian(CNR) research project No. 20-58-7802(RFBR).
\\



\end{document}